\documentstyle[aps,prb]{revtex}
\begin{document}

\title
{ 
Ab initio optical properties of
Si(100)
}

\author{
Maurizia Palummo,
Giovanni Onida,
Rodolfo Del Sole,
}
\address{Istituto Nazionale per la Fisica della Materia -
 Dipartimento di Fisica dell' Universit\'a di Roma Tor Vergata\\
    Via della Ricerca Scientifica, I--00133 Roma,
   Italy.}

\author{Bernardo S. Mendoza}
\address{Centro de Investigation en Optica, A.C. Le\'on,
Guanajuato, Mexico.}

\date{\today}

\maketitle
\begin{abstract}
We compute the linear optical properties of different 
reconstructions of 
the clean and hydrogenated Si(100) surface
within  DFT-LDA, using norm conserving pseudopotentials.
The equilibrium atomic geometries of the  
surfaces, determined from self-consistent total energy calculations 
within the Car-Parrinello scheme, strongly influence 
Reflectance
Anisotropy Spectra (RAS), showing differences
 between the p($2\times 2$) and c($4\times 2$) 
reconstructions.
The Differential 
Reflectivity spectrum 
for the c($4\times 2$) reconstruction shows a positive peak at 
$\hbar\omega < 1
eV$, 
in agreement
with experimental results.  
\end {abstract}
\section{Introduction}

The optical spectroscopy of surfaces is becoming more and more 
popular
as a non destructive and versatile tool of surface analysis. At 
variance
with electron spectroscopies, it does not require ultra-high 
vacuum
conditions, so that it can be used to monitor the growth of 
epilayers in
Molecular Beam Epitaxy and Metal Organic Vapour Deposition 
techniques.
However, the full potentiality of this kind of techniques 
can be exploited only by a strong
interaction of experimental and theoretical work,
due to the difficulty to interpret the spectra.

The dream of theorists is to compute realistic surface optical 
properties
for useful comparison with 
experiments, given only the atomic species and
positions.
Until now most of the calculations were  
 based on the semi-empirical tight-binding (TB) scheme,
which, in spite of its successes,\cite{ANATOLY} is not completely
satisfactory from 
the theoretical point of view;    
in fact, it is not selfconsistent, and needs some
experimental inputs to determine the electron bands and the 
optical
properties. 
Only in the last years,
due to the growing development of computational methods 
and
computer performances, calculations of the optical
properties within a first-principle approach started to
appear.\cite{MANGHI,KIPP,PULCI}
In principle, the correct procedure should be the 
calculation of the electronic states in a many body approach, 
including
self-energy effects, and taking into account  
the electron-hole interaction and local-field effects to obtain  
the dielectric function. Actually, 
the state of the art is 
the determination of the single particle electron states using
the Density Functional Theory (DFT)
scheme\cite{DFT} in the Local Density Approximation.\cite{LDA}

The Si(100) surface is extensively studied both from the 
experimental and
theoretical points 
of view, due to its technological importance.
Now, it is generally accepted
that its atomic structure is characterized 
by the presence of asymmetric buckled 
surface dimers along the $[01\overline 1]$ direction,
which implies at least a  
($2\times 1$) reconstruction.
Two other energetically competing (favored) reconstructions are 
possible, within the same top-atom bonding characteristic. 
The alternation of dimer buckling along 
the $[011]$ dimer rows leads to a p($2\times 2$) reconstruction, 
while
the alternation of buckling angles also along the direction 
perpendicular to the rows leads to the c($4\times 2$) 
phase.\cite{NORTRUP}

The actual geometry  of the surface  at room and low temperature 
is still controversial. 
The observed filled and empty surface states 
at room temperature were reasonably referred to c($4\times 2$)
domains present in a ($2\times 1$) disordered 
structure.\cite{JOHA1,JOHA2}
Moreover  
a phase transition from the ordered c($4\times 2$) to a 
disordered ($2\times 1$)
reconstruction,
with small c($4\times 2$) and p($2\times 2$) domains, has been 
observed
with LEED at 200 K.\cite{TABATA}
Recently Yokoyama et al.\cite{YOKO} showed with STM 
measurements
the suppressive influence of steps on
the phase transition leading to a long-range ordered c($4\times 
2$) structure.
From {\it ab-initio} molecular dynamics simulations
Shkrebtii et al.\cite{ROSA}
established that the room temperature Si(100) corresponds 
to a mixture of the c($4\times 2$) and p($2\times 2$) geometries, 
together with
instantaneous 
symmetric-like ``twisted dimers'' configurations.
At room temperature
Scanning Tunneling Microscopy (STM)  images show a
symmetric configuration,
due to the quick flipping of dimers between their two possible 
orientations.\cite{KUBO}
Recently Shigekawa et al.\cite{SHIGE}
showed that such flipping and the appearance of symmetric dimers in STM is 
connected 
to a migrating phase defect at domain boundaries.

In the last years different experimental 
studies, with Reflectance Anisotropy Spectroscopy (RAS)
and Surface Differential Reflectivity 
(SDR) techniques, have been performed
on Si(100), both on clean and adsorbate-covered surfaces. Several 
theoretical 
calculations are available, both based on tight-binding \cite{ANATOLY} and  
first-
principles \cite{KIPP,KRESS,GAVRI},
for the 2x1 and 2x2 reconstructions, but none is
available for the most stable c(4x2) phase.
For this reason, and also for the difficulty of preparing
monodomain step-free surfaces,
 not all features in the optical spectra  
are well understood.
The agreement between the theoretical
spectra obtained with the TB scheme and those computed
within the DFT-LDA
is not always satisfactory, confirming that the study of the optical 
properties
of this surface is not a simple matter.
From the experimental point of view, Shioda and van der Weide\cite{SHIODA}
and Jaloviar et al.\cite{LAGALLY}
pointed out recently the importance of a correct preparation of 
the clean 
surface 
(a terrace width about 1000 times larger than that obtained
on vicinal surfaces, typically of 40-160 $\AA$, can be obtained 
with
the
techniques employed in Refs. \ \onlinecite{SHIODA} and \ \onlinecite{LAGALLY})
in order to minimize spurious effects due to the steps.
In this way, the experiments are better suited to be compared 
with the
theoretical results.

The purpose of this work is to produce a thourough {\it ab-initio}
calculation of the optical properties of this surface, including
the ground-state c(4x2) reconstruction, to be used to interpret
RA spectra. We also calculate the optical properties of
the hydrogen-covered surface and use them to determine SDR spectra.

The article is organized as follows: in Sec. \ref{theo} we review the 
theory
for the calculation of the optical spectra. In Sec. \ref{met}, the 
method of
calculation is explained and in Sec. \ref{optics} the optical spectra 
for
the various surfaces are presented and compared with the 
experiments.
Finally, the conclusions are given in Sec. \ref{con}.

\section{Theory}\label{theo}

The surface contribution to the reflectivity is defined as the 
deviation of  
the 
reflectance with respect to the Fresnel formula, 
which is valid for an abrupt surface.  
The general formulas for the reflection coefficient 
of $s$ and $p$ radiation can be found
solving the light-propagating equations at surfaces, when 
inhomogeneity 
and anisotropy at the surface are 
fully taken into account (see Ref.\ \onlinecite{DELSOLE} for a 
review).
For normally incident light  
the correction to the Fresnel formula for the reflectivity 
is\cite{MANGHI}
\begin{equation}
\frac{\Delta R_{i}(\omega)}{R_0(\omega)} 
= 
\frac{4\omega}{c}Im\frac{\Delta\epsilon_{ii}(\omega)}{\epsilon_b
(\omega) - 1
},
\end{equation}
where $\epsilon_b$ is the bulk dielectric function, $R_0$ is the 
standard 
Fresnel reflectivity, and 
the subscript $i$ refers to the direction of light polarization. 
$\Delta\epsilon_{ii}$ is directly related to the macroscopic 
dielectric 
tensor 
$\epsilon_{ij}$ of a semi-infinite solid,\cite{FIORINO} and is
dimensionally a length.
In a repeated slab geometry, introduced in order  to simulate the 
real 
surface, $\Delta\epsilon_{ii}$ 
is given by 
\begin{equation}
\Delta\epsilon_{ii} = d[ 1 + 4\pi\alpha^{hs}_{ii} - 
\epsilon_b(\omega)],
\end{equation}
where $d$ is half of the slab thickness and $\alpha^{hs}_{ii}$ is 
the half-slab
polarizability.
For this geometry, the change of reflectivity with respect to the 
Fresnel 
formula reduces to :
\begin{equation}
\frac{\Delta R_{i}(\omega)}{R_0(\omega)}
= \frac{4\omega d}{c}Im \frac{4 
\pi\alpha^{hs}_{ii}(\omega)}{\epsilon_b(\omega) - 1}.
\end{equation}

The imaginary part of $\alpha^{hs}_{ii}$ in the single particle 
scheme 
adopted here can be 
related to the transition probabilities between slab eigenstates 
\cite{MANGHI}
\begin{equation}
\Im m[\alpha^{hs}_{ii}(\omega)] = \frac{\pi e^2}{m^2\omega^2 Ad}
\sum_{\vec{k},v,c}|p^i_{v,c}(\vec{k})|^2 \delta (E_c(\vec{k})-
E_v(\vec{k})- \hbar\omega),
\end{equation}
where $p^i_{v,c}(\vec{k})$ denotes the matrix element of the i-
component of
the momentum operator between valence states ($v$)  and 
conduction states
($c$), and $A$ is the area of the sample surface.
The k-space integration is written as a sum over $\vec{k}$ vectors 
in 
the two dimensional Brillouin zone. 
The real parts of the half-slab polarizability and bulk dielectric 
function 
are obtained via the Kramers-Kronig transform.
In the presence of a non-local potential,   
the momentum operator in (4), describing the coupling of 
electrons 
with the radiation, should be replaced by the velocity operator.
\cite{STARACE}
However, Pulci et al.\cite{PULCI} have shown that
to neglect this effect 
has a negligible influence on the RA.

Using Eq. (3) and Eq. (4) it is then
possible to compute  RAS and SDR spectra.
In particular, the
RA measures the difference in the reflectivity,  
as function of photon energy, for light polarized in two orthogonal 
directions in the specimen surface, i.e.
\begin{equation}
\frac{\Delta R}{R_0} = \frac{\Delta R_y - \Delta R_x}{R_0}.
\end{equation}
On the other hand, SDR measures the difference of the reflectance
(with unpolarized or polarized light)
between the clean and the 
covered surface: 
\begin{equation}
\frac{\Delta R}{R} = \frac{\Delta R_{\text{clean}} - \Delta 
R_{\text{cov}} }
{R_{\text{0}}}.
\end{equation}

\section{Method}\label{met}

The electronic wavefunctions and eigenvalues are obtained 
from a 
self-consistent calculation, 
in the standard DFT-LDA scheme,\cite{DFT}
using a plane-waves
basis set and the Ceperley-Alder exchange-correlation 
potential as parametrized by Perdew and Zunger.\cite{LDA}
The electron-ion interactions are treated by norm-conserving, 
fully 
separable ab-initio pseudopotentials\cite{BHS,KB}, and
a kinetic energy cutoff of 15 Rydberg is used. The crystal
with its 
surface is simulated
by a repeated slab; 
12 atomic layers and a vacuum region of   
4 empty layers are needed to obtain a good convergence of the 
optical spectra of Si(100). 
For each structure considered the atomic positions correspond 
to the fully relaxed configuration 
obtained by Car-Parrinello molecular dynamics runs. 
Three different reconstructions, ($2\times 1$), p($2\times 2$) and 
c($4\times 2$), were
considered.
In Tab. 1 
we report the structural parameters obtained for the various 
reconstructions 
of the clean surface. 
In order to obtain SDR spectra,
calculations were performed also for a passivated surface
 (two Hydrogens per surface Si atom-- yielding a bulk-like terminated (1x1) 
surface without any dimers),
using the same ingredients as for the corresponding 
clean surfaces.  
In the calculation of the optical spectra, we verified that the use of
a reduced kinetic energy cut-off (10 Ry. instead of 15) does not
change the resulting spectra appreciably. The gain in computational
cost was then exploited to perform very extended convergence tests
with respect to the k--points sampling used. We found 
that for the ($2\times 1$) and hydrogenated surfaces
a 64 k-points set 
is necessary and sufficient to obtain converged spectra in the range
0-5 eV, 
while for the c($4\times 2$) and p($2\times 2$) reconstructions, 
due to the smaller BZ,
32 k points are enough. 
For each surface, we compute equation (4) using a number of 
empty 
states within 1 Ry from 
the conduction band minimum.

In order to compare the theoretical spectra with the experimental 
curves,  
we apply an upward  rigid shift $\Delta$ = 0.5 eV to the 
conduction 
bands (in agreement with the results of previous 
GW calculations
\cite{NORTRUP,KRUGER} on this surface)
due to the well known underestimation 
of the gaps in the  DFT-LDA method. 
The momentum matrix elements are renormalized 
with the factor $(E_c -E_v + \Delta)/(E_c-E_v)$,
according to the recipe of Del Sole and Girlanda.\cite{GIRLA}

\section{Optical spectra}\label{optics}

We obtain band structures 
in good agreement with other, well converged DFT-LDA 
calculations appeared in the
literature,\cite{KRUGER,RAMSTAD}
in particular with those of Ref. \onlinecite{RAMSTAD}.
For the ($2\times 1$) reconstruction, the $\pi$ and $\pi^*$ surface 
states show a rather
strong 
dispersion along the dimer rows ($J-K$, $J'-\Gamma$ directions). This 
implies a 
relevant interaction between adjacent dimers in the same row,
while the flatter dispersion along the dimer direction (i.e.,
perpendicularly to the row) indicates a 
small 
interaction of dimers belonging to different rows.  
The  c($4\times 2$) and  p($2\times 2$) reconstructions,
in contrast with the ($2\times 1$) case, show
two occupied and two empty dangling-bond surface bands,  
in agreement with ARUPS experiments.\cite{JOHA1}
The dispersion of the unoccupied $\pi^*$ surface state of the 
c($4\times 2$)
reconstruction 
is in good agreement with recent standing wave patterns 
measurements, 
obtained with STM at low temperature \cite{YOKO2}.

In Fig. 1 we show  
the RA spectra obtained for the ($2\times 1$)
surface using three different geometries: the first
 with symmetric dimers, the second with a
buckling of 0.56 $\AA$  (similar to Kipp's calculation\cite{KIPP})
 and the last of 0.71 $\AA$  (which corresponds to our fully
converged geometry at the cut-off of 15 Ry).
The changes in the theoretical
results are 
evident: the first low energy peak,  
essentially due to dimer surface-states, 
shifts from 1 eV to 1.4 eV
(due to an opening of the gaps) and
changes its sign,  in going from the symmetric to the asymmetric 
case. The same peak 
increases in strength and shifts slightly to higher energy (1.6 eV)
with the increase of the buckling.
On the other hand, the positive peak at 4.3 eV, 
which is always present in the experimental spectra, 
reduces with the increase of the buckling.  
\par\noindent
\par\noindent

Fig. 2 shows the RAS obtained for the c($4\times 2$) 
reconstruction.
Our results for the p($2\times 2$) and ($2\times 1$) spectra are 
also included. The latter  are
consistent
with the results obtained by Kipp et al.,\cite{KIPP} although not
identical due 
to the different equilibrium geometries at which the optical 
spectra 
have 
been
calculated.
The c($4\times 2$) 
and p($2\times 2$) spectra
show however a larger anisotropy on the low frequency side, due 
to 
transitions across the folded dangling-bond bands, 
while the ($2\times 1$) reconstruction has a single negative peak in 
this 
range, at about 1.4 eV.
Differences are present expecially below $\hbar\omega = 1.5 eV$ 
between the c($4\times 2$) and p($2\times 2$) reconstructions, 
 confirming the strong dependence 
of the spectra upon the details of the reconstruction, 
in particular upon the ordering of the buckled dimers along and 
perpendicularly to the rows. 
The comparison of the overall shape with the measurements 
obtained 
by Shioda and van der Weide\cite{SHIODA}
is generally good, showing that the various features of the 
experimental
spectra are due essentially to the different 
reconstructions present in the specimens.\cite{NOTA} 

The interpretation of the experimental spectra is complicated by 
the disordered nature of the dimer
buckling at room temperature, and by spurious effects due to 
the 
steps, 
or defects, 
that cannot be completely eliminated. 
The negative peak close to 3.7 eV can be explained by any of the three
structures considered in Fig. 2, while the 3 eV structure implies the presence
of (2x1) domains at room temperature.
Especially for the peak at about 4.3 eV, the comparison 
with the experiment is not completely clear. 
Cole  et al.\cite{COLE}
found 
that its 
intensity is different in different samples, 
 and decreases almost linearly while increasing the temperature.
Shioda and van der Weide\cite{SHIODA} found a similar behavior
by decreasing the  
miscut angle of the vicinal surfaces, which corresponds to
reducing the majority domain coverage (or, in other words, 
the anisotropy of the surface) . 
Yasuda et al.\cite{YASUDA}, from the analysis of their RT
spectra,
concluded that this peak
is essentially originated by a step induced dicroism; however, the
opposite conclusion has been reached by Jaloviar et al.\cite{LAGALLY}, who
singled out the contribution of the steps to the RA.  
In our spectra a peak  
in the energy region around 4.3 eV is present for the p($2\times 
2$) and c($4x2$)
reconstructions, in agreement with the observed reduction with 
increasing temperature.
On the other hand, as shown in Fig. 1,  a peak at the same 
energy is
present also 
in the ($2\times 1$) case, and its intensity  depends on the 
dimer buckling. Hence, we conclude, in agreement with Refs. \ \onlinecite{SHIODA}
and \ \onlinecite{LAGALLY}, that this structure does not arise
from the steps.

SDR experiments on the clean Si(100) surface
for $\hbar\omega < 1 eV$
were 
carried out in 1983 by Chabal et al.\cite{CHABAL}
using the multiple internal total reflection arrangement.
In 1980 Wierenga at al.\cite{WIERE1},  and more recently Keim 
et al.  \cite{KEIM}, measured SDR   
using normal-incidence external reflection.
Hydrogen, oxygen, or water were used in the experiments 
in order to saturate the dangling bonds (only oxygen was used 
in the last one).
In Fig. 3 we report theoretical SDR
spectra  calculated using, as the clean surface contribution,
the c($4\times 2$),
the p($2\times 2$) and the ($2\times 1$) reconstructions, 
respectively. The ($1\times 1$) H-covered surface is used as the reference
 surface. The
experimental results of 
Ref. \onlinecite{CHABAL} and \onlinecite{WIERE1} are also 
shown.
The positive peak at about  1.5 eV, present in 
the experimental spectra, 
is confirmed
by our theoretical analysis. 
The first peak  below 1 eV found by Chabal et al. 
\cite{CHABAL} appears only when
the reflectivity of the clean c($4\times 2$) 
reconstruction
is used in eqs. 4 and 6, although a shoulder is also present 
for the p($2\times 2$) reconstruction.
(A peak at the same energy
is visible also in EELS spectra.\cite{EELS})
Experimentally, the intensity of this peak is found to be weakly 
dependent on the temperature
from 40 K to room temperature.
Different 
explanations of this structure have been proposed in the past:  
in EELS experiments it was interpreted as a direct
transition at $\Gamma$ from the bulk top valence to an empty 
surface state, 
while in SDR experiments it was explained as an indirect 
transition.
Instead, in Ref.  
\onlinecite{ANATOLY} it was interpreted in terms of transitions 
across the
dangling-bonds of flipping dimers, when these are 
instantaneously 
unbuckled: 
in this position indeed, the energy separation has the right 
value to produce a peak just below 1 eV (see the dotted curve in 
Fig. 1). 
Although this explanation cannot be ruled out,   
we may now argue that the SDR structure below 1 eV 
could be due to the presence of c($4\times 2$) and 
p($2\times 2$) 
domains, where it
 is essentially due to transitions from bulk
states near the top valence to unoccupied surfaces states. 
The
 small intensity of the experimental peak may be due to the small 
fraction of 
c($4\times 2$) domains, caused by thermal disorder as well as 
 by the presence of defects or 
steps,
which can suppress the 
long-range ordered structures.

At higher energies, a peak at about 2.5 eV is reported 
in ref. \onlinecite{WIERE1}
while two peaks at 2.90 and 3.95 eV 
are present in the experimental
spectra of Ref. \onlinecite{KEIM}
when the surface is exposed to NO$_2$, and a further small peak
appears  
at 3.1 eV when molecular oxygen is used.  
The authors of Ref. \onlinecite{KEIM}
argue that the peaks at 2.9 and 3.95 eV
are related to transitions between 
surface states; instead,  
in our theoretical analysis for
all the reconstructions studied, all spectral features
above 2 eV are essentially due to surface-bulk, bulk-surface
and bulk-bulk transitions.
\par
\par
The agreement between our DFT-LDA RA spectra and 
those obtained in the TB method
is limited, even if the same
geometry is used. We show in Fig. 4a the RA calculated according
to the two methods for the 2x1 reconstruction, using a
 buckling of 0.56 \AA (as for the full-line curve in Fig. 1), which is very
close to that of Ref. \ \onlinecite{ANATOLY}.  
As previously found,\cite{KIPP,KRESS}
the sign of the RAS at about 1.5 eV, due to transitions
across surface states, is
opposite to that obtained with the nearest-neighbors TB 
calculations. 
The latter coincides with that  
of a naive picture of the dimers
considered as isolated Si$_2$ molecules.
This can be explained on the basis of the omission, in the TB
approach, of the direct
interaction
between atoms belonging to adjacent dimers.
In fact, the explanation of this apparently 
paradoxical sign of the reflection anisotropy  
can be found looking at the surface band 
structure:\cite{RAMSTAD}
 the dispersion of the surface-localized $\pi$ and $\pi^*$ bands 
along the row
direction (i.e., perpendicularly to the dimers) is comparable to the  $\pi$-
$\pi^*$ separation. This means that the interaction between 
adjacent dimers is about as large as the interaction between the 
$p_z$ orbitals of the two atoms in a single dimer.
Hence, the correct picture of these surfaces is that of interacting 
chains of dimers, oriented in the direction perpendicular to the 
dimers themselves, and separated by large valleys through which 
the dimers 
interact very weakly. 
Recently Gavrilenko and Pollack performed theoretical 
calculations
both with the DFT-LDA and the TB schemes.\cite{GAVRI}
However, while their TB results agree with our calculations, their 
DFT-
LDA spectra,
which are obtained with some kind of ``ad-hoc'' treatment of the 
self-
energy effects, 
appear to be at variance with both our results and the calculations 
of 
ref. \onlinecite{KIPP}, and do not describe 
correctly the experimental spectra. 

In Fig. 4 (b) we compare the {\it unpolarized}
reflectivity spectra obtained in our TB and LDA 
calculations, 
using the same geometry (buckling of 0.56 $\AA$). In this
case, our ab--initio calculation of the SDR 
spectra 
basically agree with the	TB results and with
previously published data \cite{ANATOLY}, 
confirming that once averaged over the polarizations the spectra 
depend less critically 
on the theoretical scheme used. In particular, the occurrence of the
(experimentally measured) first peak at about 1.4 eV 
as a consequence
of dimer buckling, which was the main point of Ref. \onlinecite{ANATOLY}, is confirmed.
In the case of unbuckled dimers, in fact,
 the first SDR peak occurs at 0.9 eV, where
the first RAS structure in Fig.1 occurs.
 
\section{Conclusions}\label{con}

We have carried out {\it ab-initio} calculations of the optical 
properties 
of the Si(100) surface within DFT-LDA. For the first time this 
approach 
has been applied to the c($4\times 2$) clean surface and to the 
($1\times 1$) hydrogenated
surface. Quite surprisingly, the optical properties of the  
c($4\times 2$)
reconstructed surface are qualitatively different from those of the 
p($2\times 2$) phase on the low-frequency side.
A qualitative understanding of RAS and SDR measurements is 
obtained. 
The structure appearing below 1 eV in the SDR spectrum, not well 
understood so far, is naturally explained in terms of the 
occurrence of 
c($4\times 2$) domains.

Our {\it ab-initio} DFT-LDA calculations allow us to check the 
reliability 
of the TB approach. SDR TB
spectra are in qualitative agreement with DFT-LDA spectra,
while some discrepancy appears in the case of
 RAS spectra, which are more 
sensitive to 
the details of the local geometry and require more accurate 
calculations. 
The omission of second-neighbor interactions in the $sp^3s^*$ 
scheme
used in our and previous TB calculations is mostly responsible for 
this
discrepancy. Calculations using a second-neighbours parametrization
scheme should correct this shortcoming of the TB approach. 

\section{Acknowledgments}

 We are grateful to  S. Goedecker for providing us an efficient
code  for Fast Fourier Transforms.\cite{goedecker}
 This work was supported in part by the European Community
 programme ``Human Capital and Mobility'' (Contract
 No. ERB CHRX CT930337), and by the INFM Parallel Computing Initiative.
 Calculations were performed at CINECA (Interuniversity Consortium of the
 Northeastern Italy for Automatic Computing).
B.S.M. thanks partial support of CONACyT-M\'exico (26651-E) and
CNR-Italy through a collaboration program.

\narrowtext
\begin{table}
\caption{Calculated buckling, and dimer
length, 
 for the ($2\times 1$),c($4\times 2$),
 and p($2\times 2$) Si(100) reconstructions. In parenthesis,
the value obtained if the self--consistent geometry optimization
is performed at an unconverged kinetic energy cutoff (10 Ry.)}
\begin{tabular}{  l  l  l }
Reconstruction & Dimer buckling ($\AA$) & Dimer length ($\AA$)\\
\hline 
($2\times 1$) & 0.71(0.56) & 2.27(2.34) \\
c($4\times 2$)  &  0.75  & 2.29  \\
p($2\times 2$)  &  0.70  & 2.28  \\
\end{tabular}
\label{table1}
\end{table}
\widetext
\begin{figure}
\caption{ Theoretical DFT-LDA RA spectra
(a scissor operator of 0.5 eV is applied 
as discussed in the text) for the 
$2\times 1$ clean surface, as a function of dimer buckling. 
The buckling of 0.0$\AA$ (dotted line) corresponds to
symmetric dimers. The reflectance anisotropy is defined as in equation (5),
 where
y (x) is the direction parallel (perpendicular) to the dimers.
}
\label{fig1}
\end{figure}

\begin{figure}
\caption{  
RA (defined as in Fig. 1) theoretical spectra  obtained with the different 
Si(100) reconstructions considered: full line: ($2\times 1$); dashed 
line:
p($2\times 2$);
dotted line: c($4\times 2$). The experimental 
results,\protect\cite{SHIODA}
scaled up by a factor of 5,
are also shown.  }
\label{fig2}
\end{figure}

\begin{figure}
\caption{ 
SDR theoretical spectra obtained
considering different reconstructions for the clean surface: 
full line: ($2\times 1$);
dashed line: p($2\times 2$); dotted line: c($4\times 2$).
The hydrogenated surface is (1x1) with 2 monolayers of H.
The experimental results from Refs.
{\protect\onlinecite{CHABAL}}
 and {\protect\onlinecite{WIERE1}} are  shown by squares 
and dots, respectively.}
\label{fig3}
\end{figure}

\begin{figure}
\caption{ 
Comparison of plane-wave and TB results for the reflectance 
anisotropy 
(defined as in Fig. 1, panel a) 
and for the unpolarized reflectivity (panel b), calculated  for Si(100) ($2\times 1$).}
\label{fig4}
\end{figure}


\begin{references}
\bibitem{NOTA} The experimental RAS data of references {\protect \onlinecite{SHIODA}} and
{\protect\onlinecite{LAGALLY}} are similar. We quote in Fig. 2 those of Ref.{\protect\onlinecite{SHIODA} }
because they include the frequency range from 1 to 2 eV, which is especially
interesting for surface-state transitions.
\end{references}
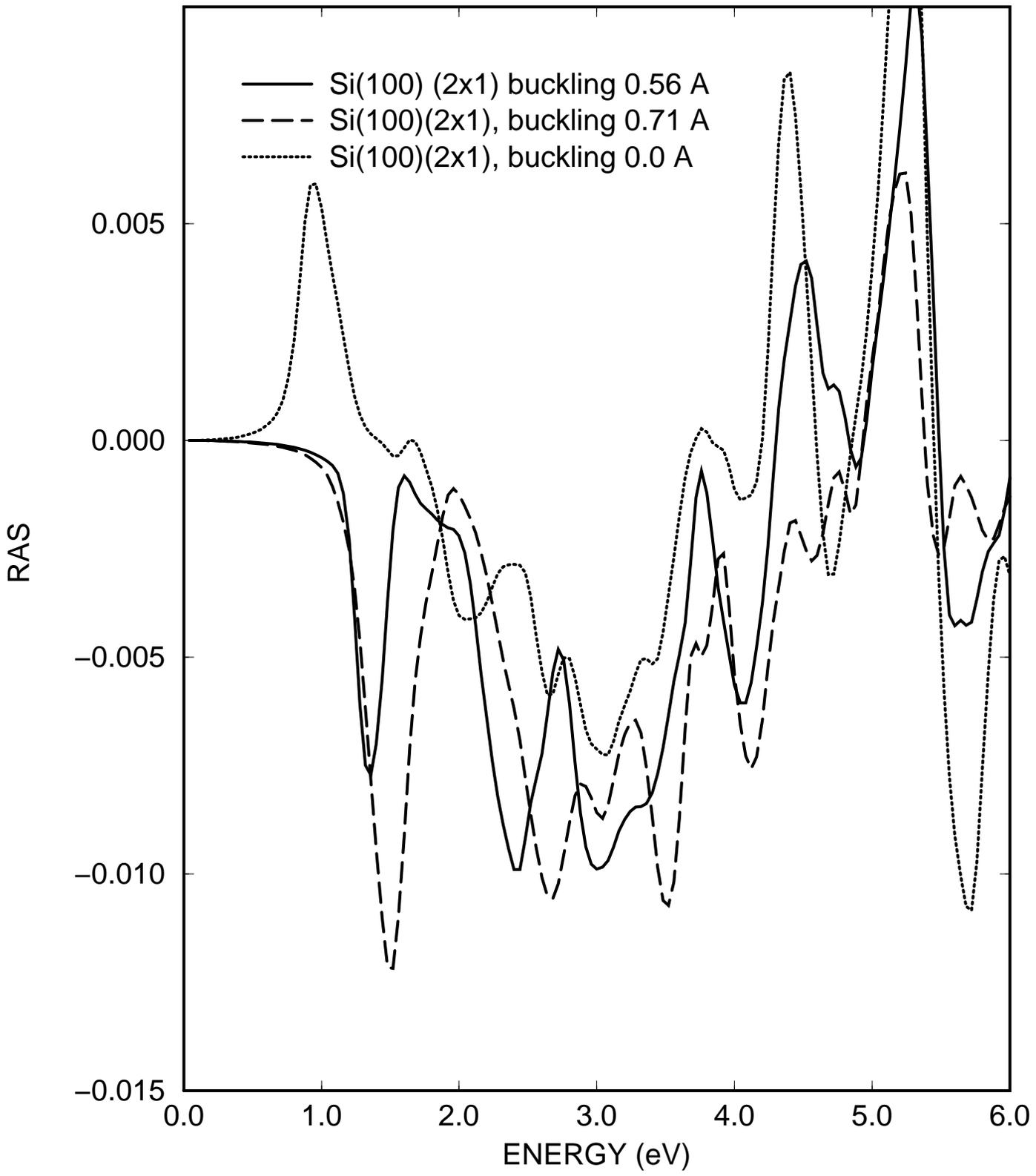
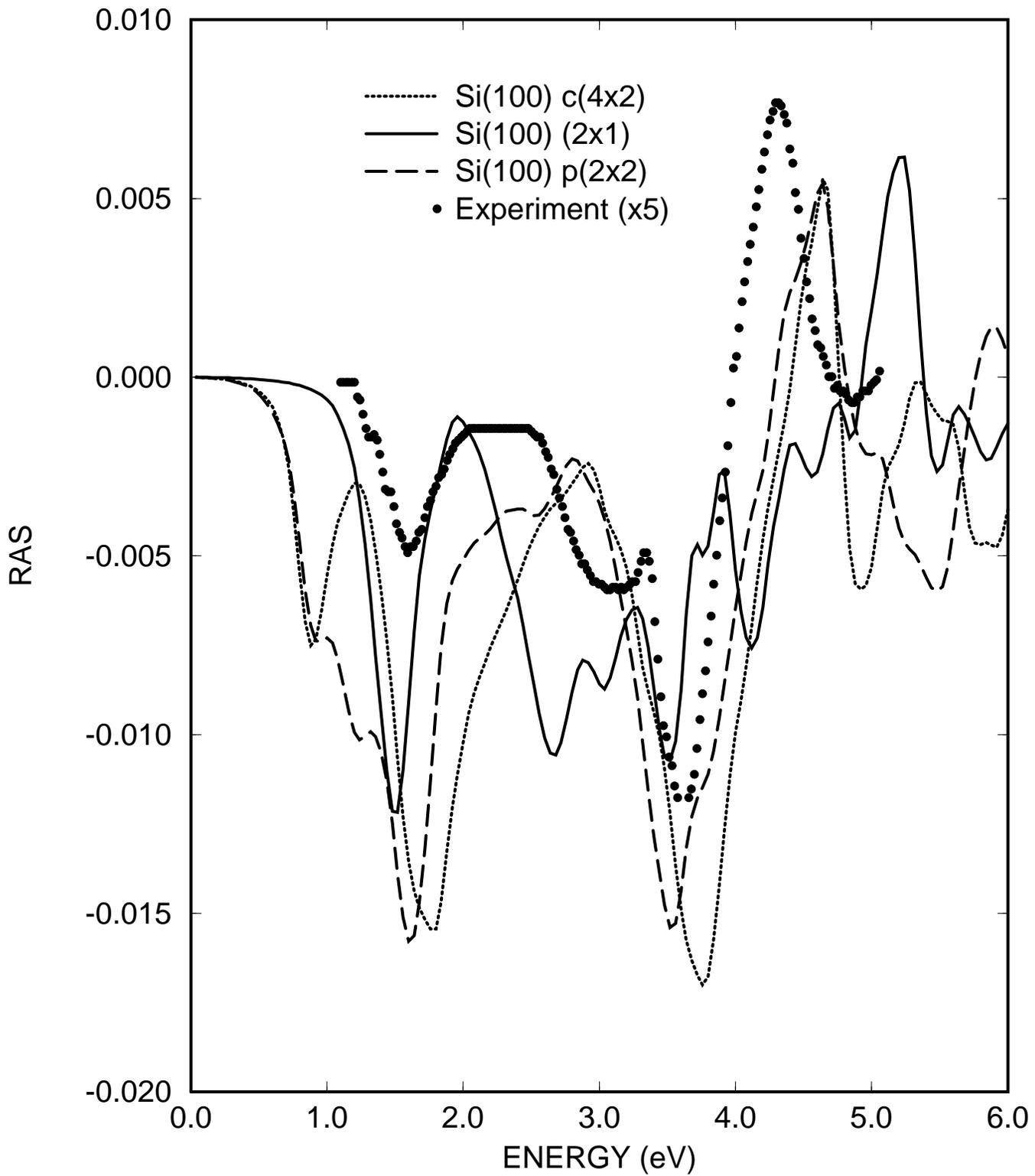
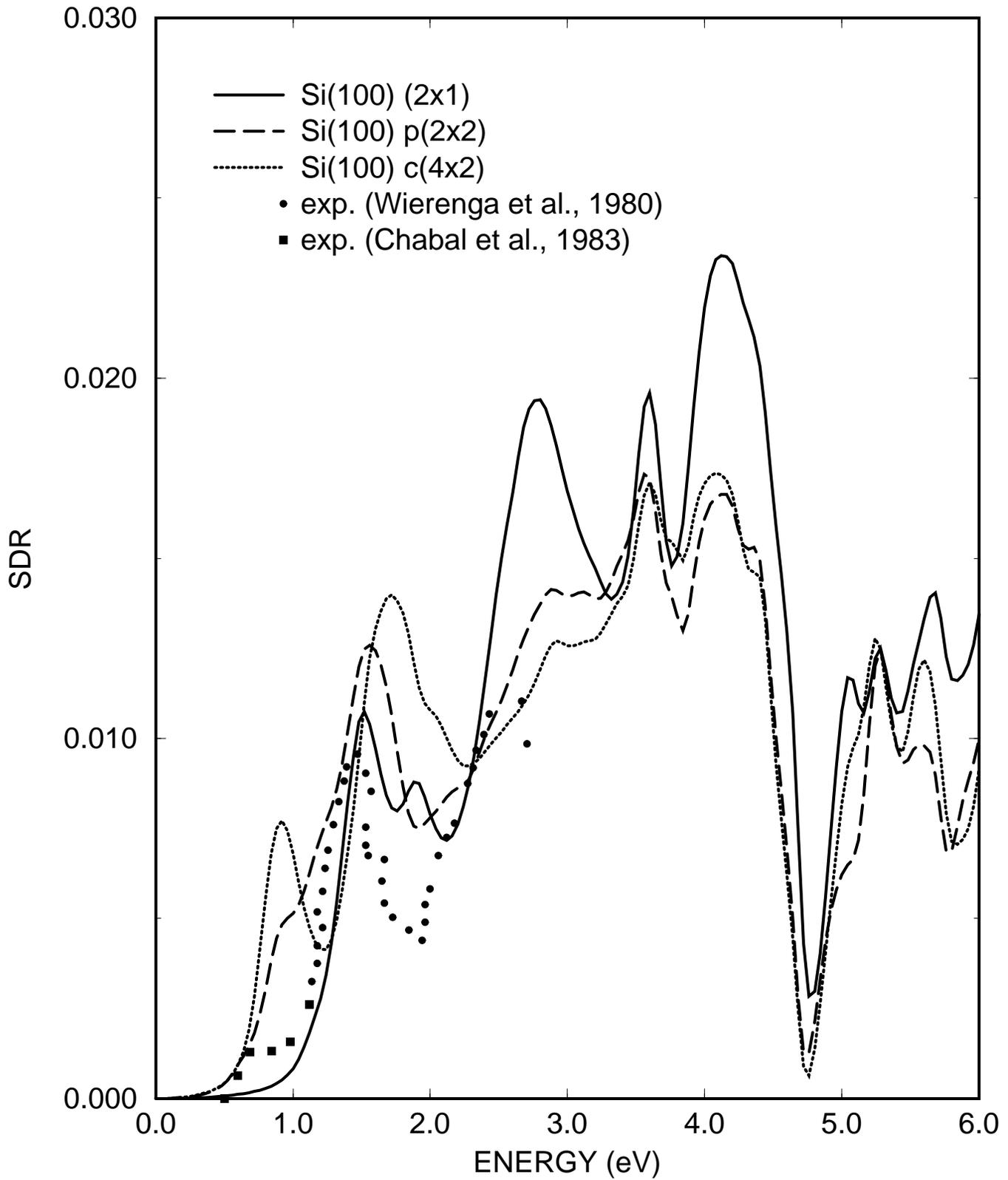
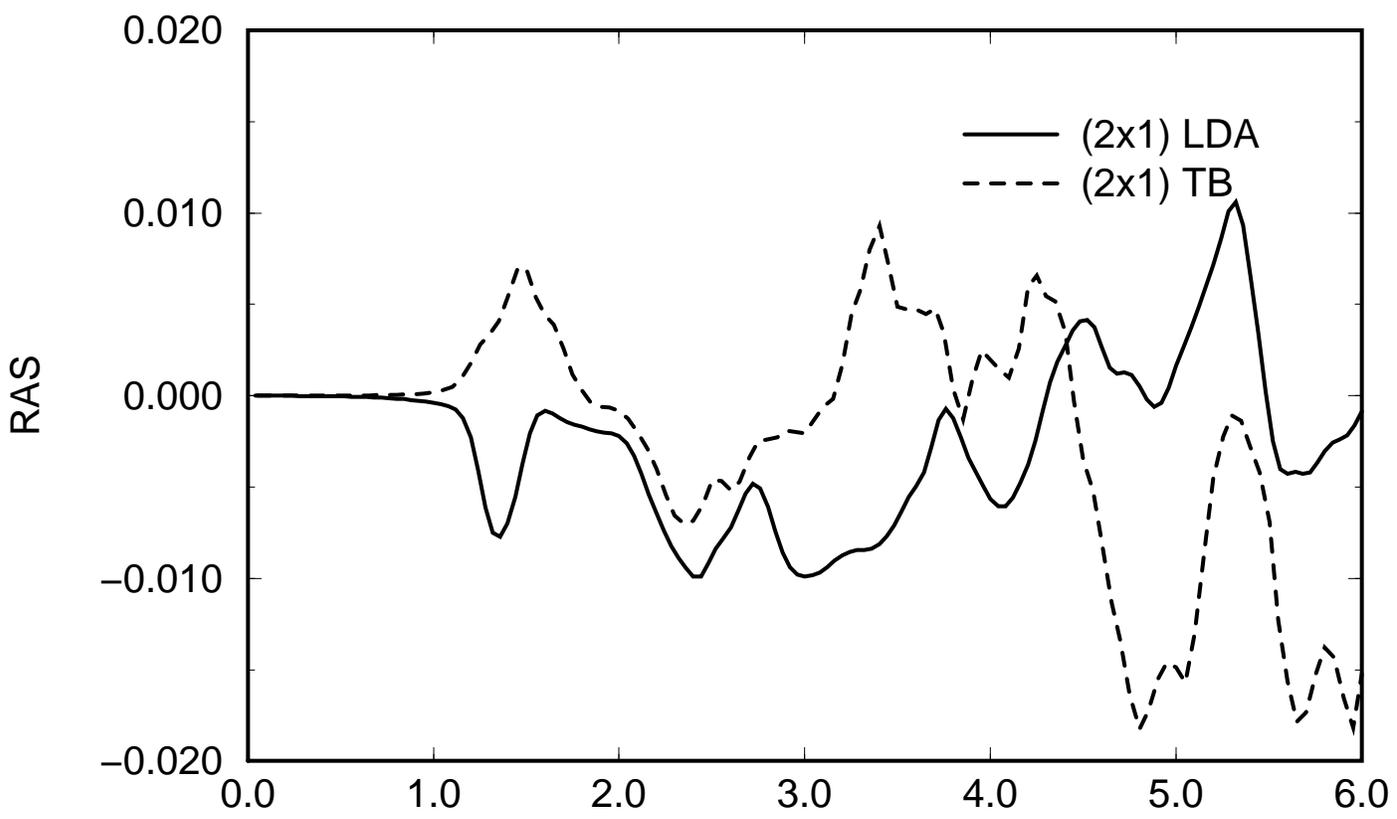
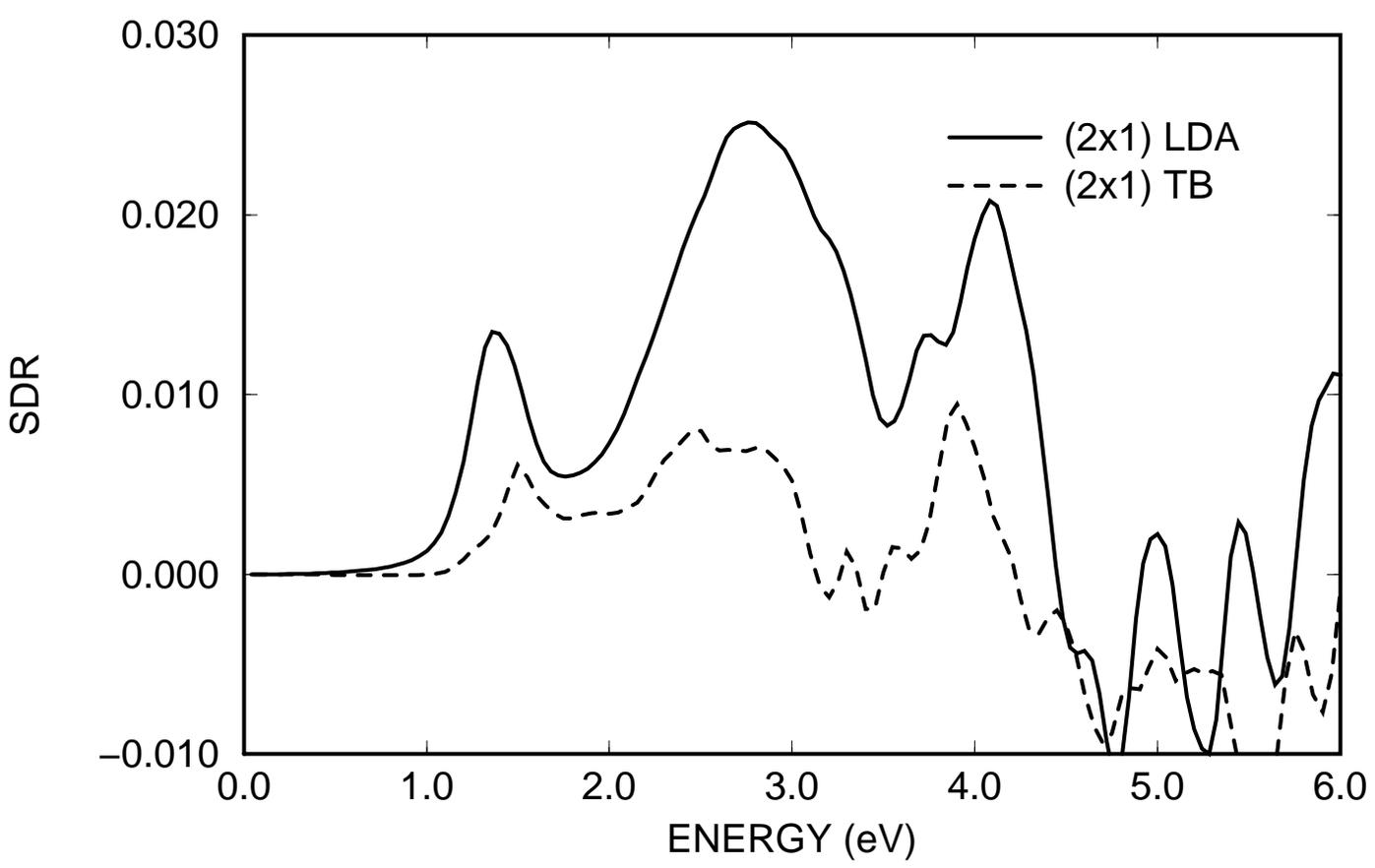
\end{document}